# Monolithic Integration of AlGaAs Distributed Bragg Reflectors on Virtual Ge Substrates via Aspect Ratio Trapping

YIHENG LIN,[1] WEI SHI,[2] JIZHONG LI,[3] TING-CHANG CHANG,[1,4] JI-SOO PARK,[3] JENNIFER HYDRICK,[3] ZIGANG DUAN,[5] MARK GREENBERG,[2] JAMES G. FIORENZA,[3] LUKAS CHROSTOWSKI[2] AND GUANGRUI (MAGGIE) XIA[1,*]

[1] *Department of Materials Engineering, the University of British Columbia, Vancouver BC, V6T 1Z4 Canada*
[2] *Department of Electrical and Computer Engineering, University of British Columbia, Vancouve, BC, V6T 1Z4 Canada*
[3] *AmberWave Systems Corp., NH 03079, USA*
[4] *Department of Physics, National Sun Yat-Sen University, Kaohsiung 804, Taiwan, ROC*
[5] *Key Laboratory of Optoelectronic Devices and Systems, Shenzhen University, Shenzhen 518060, China.*
*[*]gxia@mail.ubc.ca*

**Abstract:** High quality $Al_xGa_{1-x}As$ distributed Bragg reflectors (DBRs) were successfully monolithically grown on on-axis Si (100) substrates via a Ge layer formed by aspect ratio trapping (ART) technique. The GaAs/ART-Ge/Si-based DBRs have reflectivity spectra comparable to those grown on conventional bulk off-cut GaAs substrates and have smooth morphology, and quite reasonable periodicity and uniformity. Antiphase domain formation is significantly reduced in GaAs on ART-Ge/Si substrates, and etch pit density of the GaAs base layer on the ART-Ge substrates ranges from $10^5$ to $6 \times 10^6$ cm$^{-2}$. These results paved the way for future VCSEL growth and fabrication on these ART-Ge substrates and also confirm that virtual Ge substrates via ART technique are effective Si platforms for optoelectronic integrated circuits.

## References

1. A. Alduino and M. Paniccia, "Interconnects: Wiring electronics with light," Nat. Photon. **1**, 153-155 (2007).
2. R. Soref, "The Past, Present, and Future of Silicon Photonics," IEEE J. Sel. Top. Quantum Electron. **12**, 1678-1687 (2006).
3. X. Sun, J. Liu, L. Kimerling, and J. Michel, "Toward a Germanium Laser for Integrated Silicon Photonics," IEEE J. Sel. Top. Quantum Electron. **16**, 124-131 (2010).
4. K. Iga, "Surface-emitting laser-its birth and generation of new optoelectronics field," IEEE J. Sel. Top. Quantum Electron. **6**, 1201-1215 (2000).
5. H.J. Yeh and J.S. Smith, "Integration of GaAs vertical-cavity surface emitting laser on Si by substrate removal," Appl. Phys. Lett. **64**, 1466-1468 (1994).
6. Z. Zhu, F. Ejeckam, Y. Qian, Jizhi Zhang, Zhenjun Zhang, G. Christenson, and Y. Lo, "Wafer bonding technology and its applications in optoelectronic devices and materials," IEEE J. Sel. Top. Quantum Electron. **3**, 927-936 (1997).
7. M. Kwack, K. Jang, J. Joo., H. Park, J. H. Oh, J. P. Park, S. K. Kim and G. Kim, "Device characterization of the VCSEL-on-silicon as an on chip light source," Proc. of SPIE **9752**, 97521A (2016).
8. G. Kim, H. Park, J. Joo, K. Jang, M. Kwack, S. Kim, I. G. Kim, J. H. Oh, S. A. Kim, J. Park and S. Kim, "Single-chip photonic transceiver based on bulk-silicon, as a chiplevel photonic I/O platform for optical interconnects," Sci. Rep. **5**, 11329 (2015).
9. D.G. Deppe, N. Chand, J.P.V.D. Ziel, and G.J. Zydzik, "$Al_xGa_{1-x}As$-GaAs vertical-cavity surface-emitting laser grown on Si substrate," Appl. Phys. Lett. **56**, 740-742 (1990).
10. T. Egawa, Y. Murata, T. Jimbo, and M. Umeno, "Characterization of AlGaAs/GaAs vertical-cavity surface-emitting laser diode grown on Si substrate by MOCVD," Appl. Surf. Sci. **117-118** (6), 771-775 (1997).
11. V.K. Yang, M. Groenert, C.W. Leitz, A.J. Pitera, M.T. Currie, and E.A. Fitzgerald, "Crack formation in GaAs heteroepitaxial films on Si and SiGe virtual substrates," J. Appl. Phys. **93**, 3859-3865 (2003).

12. M.T. Currie, S.B. Samavedam, T.A. Langdo, C.W. Leitz, and E.A. Fitzgerald, "Controlling threading dislocation densities in Ge on Si using graded SiGe layers and chemical-mechanical polishing," Appl. Phys. Lett. **72**, 1718-1720 (1998).
13. J. Park, J. Bai, M. Curtin, B. Adekore, M. Carroll, and A. Lochtefeld, "Defect reduction of selective Ge epitaxy in trenches on Si(001) substrates using aspect ratio trapping," Appl. Phys. Lett. **90**, 052113 (2007).
14. J.-S. Park, M. Curtin, J. Bai, M. Carroll, and A. Lochtefeld, "Growth of Ge Thick Layers on Si(001) Substrates Using Reduced Pressure Chemical Vapor Deposition," Jpn. J. Appl. Phys. **45**(11), 8581–8585 (2006).
15. J. Z. Li, J.M. Hydrick, J.S. Park, J. Li, J. Bai, Z.Y. Cheng, M. Carroll, J.G. Fiorenza, A. Lochtefeld, W. Chan, and Z. Shellenbarger, "Monolithic Integration of GaAs/InGaAs Lasers on Virtual Ge Substrates via Aspect-Ratio Trapping," J. Electrochem. Soc. **156**, H574-H578 (2009).
16. Z. Wang, B. Tian, M. Pantouvaki, W. Guo, P. Absil, J. V. Campenhout, and D. V. Thourhout, "Room-temperature InP distributed feedback laser array directly grown on silicon," Nat. Photon. **9**(12), 837-842 (2015).
17. J.-S. Park, J. Bai, M. Curtin, B. Adekore, M. Carroll, A. Lochtefeld, "Defect reduction of selective Ge epitaxy in trenches on Si(001) substrates using aspect ratio trapping," Appl. Phys. Lett. **90**, 052113 (2007).
18. J. Bai, J.-S. Park, Z. Cheng, M. Curtin, B. Adekore, M. Carroll, A. Lochtefeld, "Study of the defect elimination mechanisms in aspect ratio trapping Ge growth," Appl. Phys. Lett. **90**, 101902 (2007).
19. P. Zaumseil, T. Schroeder, J. Park, J.G. Fiorenza, and A. Lochtefeld, "A complex x-ray structure characterization of Ge thin film heterostructures integrated on Si(001) by aspect ratio trapping and epitaxial lateral overgrowth selective chemical vapor deposition techniques," J. Appl. Phys. **106**, 093524 (2009).
20. M.S. Abrahams and C.J. Buiocchi, "Etching of Dislocations on the Low-Index Faces of GaAs," J. Appl. Phys., **36**, 2855-2863 (1965).

## 1. Introduction

Over the past three decades, researchers have devoted great efforts on silicon (Si) photonics to overcome the communication bottleneck of integrated circuits [1-2]. Excellent performance has been achieved so far on waveguides, modulators and detectors for short-reach optical interconnects, which use Si compatible materials (e.g. $SiO_2$, $Si_3N_4$ and SiGe) and processes. However, lasers on Si have been much more difficult to implement [3]. Vertical cavity surface emitting lasers (VCSELs), compared to edge emitting lasers, are very suitable as output devices for optical interconnects on Si platforms. The advantages of VCSELs include high-density two-dimensional array fabrication, low-cost testing and packaging, easy fiber coupling and low power consumption [4]. To integrate VCSELs with Si, hybrid bonding approaches have been studied and shown to be effective [5-8]. Compared with hybrid integration, monolithic integration has the advantages of high density, system reliability, increased functionality and the ease of using well-established low-cost Si fabrication techniques. Several studies of monolithic integration have been reported. However, the performance was insufficient for communication applications [9-10].

As VCSELs are fabricated using compound semiconductors such as $Al_xGa_{1-x}As$ and GaAs, difficult challenges exist for the monolithic integration of VCSELs on Si platforms due to material property mismatches among different materials, such as lattice constant, thermal expansion and crystal structure mismatch, which can lead to high density defects and cracking [11]. Compared to Si, the lattice constant of germanium (Ge) is much closer to those of GaAs and AlGaAs. Many techniques have been developed to utilize Ge as a buffer material to realize high quality GaAs on Si including direct Ge layer deposition, graded buffer layer, Ge condensation, and aspect ratio trapping (ART) technique [12-15]. ART technique has the advantages of a small Si to Ge transition layer thickness and high material quality, compared to other approaches, and is our choice for VCSEL integration on Si platforms. Although the substrate is on-axis (100) Si, this growth technique has off-cut characteristics due to the formation of Ge facets, which lead to antiphase domain (APD) reduction [15]. Therefore, it is compatible with standard Si processing based on non-offcut (100) Si substrates. A similar method has also been recently used to fabricate monolithic integrated InP distributed feedback lasers on Si, where optical cavities were selectively grown on Si with the assistance of the $SiO_2$ sidewalls [16]. In this case, the threading dislocations and anti-phase boundaries were confined to a less than 20-nm-thick layer to enable the optically pumped pulse laser operation.

In this paper, we report the monolithic integration of $Al_xGa_{1-x}As$ distributed Bragg reflectors (DBRs) on Si substrates via ART technique, which is a crucial step towards the true monolithic integration of VCSELs on Si.

## 2. Epitaxy growth

The schematic structure of the DBRs grown on Ge/Si ART substrate is shown in Fig. 1. The ART substrates used in this study were p-type on-axis eight inch Si (100) substrates and the growth details were described previously [14-15]. First, a thin $SiO_2$ layer was thermally grown on the Si substrate, followed by conventional photolithography and reactive ion etching to form parallel trenches and waffle patterns along [110] direction, which are designed to study the effects of oxide trench patterns on crystal quality of the subsequent epitaxial Ge and GaAs films. The oxide trenches in all patterns are 0.25 μm wide and 0.5 μm deep. The $SiO_2$ spacer width between neighboring trenches varied from 0.25 to 20 μm for the parallel trench patterns. After trench patterning and before Ge growth, the final trench height was about 480 nm.

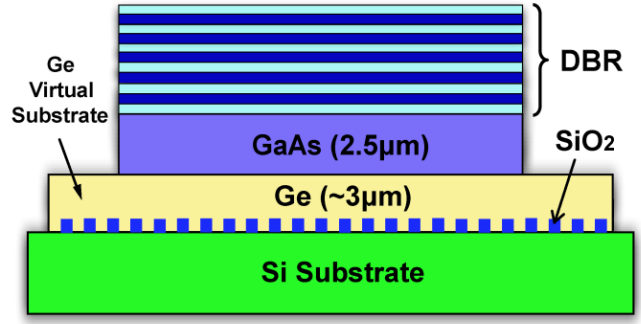

Fig. 1. Schematic structure of the DBRs grown on Ge/Si ART substrate with GaAs buffer layer.

Two-step low pressure metalorganic chemical vapor deposition (MOCVD) processes were performed in this study. First, an epitaxial Ge layer was grown on the patterned substrate under optimized growth conditions similar to that described previously [17]. Ge growth was terminated after the Ge layer slightly coalesced in the 20 μm spacer section, which corresponds to an average layer thickness of about 4 μm. Then, a CMP process was used for planarization of the Ge–$SiO_2$ composite structure. Before GaAs overgrowth, the polished Ge/Si substrate was cleaned with successive dips in $H_2O_2$ solution and 1:50 diluted HF, with deionized water rinses between steps. In the second growth step, GaAs layers were deposited in a second MOCVD reactor at a constant low pressure (70 torr) by using triethylgallium and arsine ($AsH_3$) for the buffer-layer growth and trimethylgallium, trimethylaluminum, trimethylindium, and $AsH_3$ for the upper structural-layer growth. Before the 30 nm GaAs buffer-layer growth, the wafer was baked at 600°C for 10 min under $H_2$ overpressure followed by a 2 min As coating through the introduction of $AsH_3$ overpressure. Then, the growth temperature was reduced to 400°C for 30 nm GaAs buffer-layer growth followed by a three-period GaAs (10 nm)/$Al_{0.4}Ga_{0.6}As$ (15 nm) superlattice structure grown at 600°C. Finally, an n-type 2 μm GaAs base layer was grown.

Subsequently, the GaAs/Ge-Si ART wafers were cut into pieces and transferred to a third MOCVD reactor, which only has three inch size wafer pockets. Three inch Si dummy wafers with a customized 700 μm wafer thickness and center rectangular holes were used to position the ART wafer pieces in the pockets. A 0.5 μm thick GaAs layer and a DBR superlattice were grown on the ART substrates as well as a three inch 10 degree offcut bulk GaAs substrate as the control sample. Each DBR consists of 34 periods of n-type $2\times10^{18}$ cm$^{-3}$ doped 20 nm $Al_xGa_{1-x}As$ (x: 0.12→0.9) / 50 nm $Al_{0.9}Ga_{0.1}As$ / 20 nm $Al_xGa_{1-x}As$ (x: 0.9→0.12) / 39 nm $Al_{0.12}Ga_{0.88}As$, which was designed with a center of peak reflectivity wavelength of 850 nm for 850 nm wavelength VCSELs. The DBR growth used the same process recipe of a standard commercial 850 nm VCSEL at LandMark Optoelectronics Corporation. The growth temperature was 760 °C and the gas pressure was 100 torr. The gas used in GaAs growth was $H_2$, $SiH_4$, $AsH_3$ and $Ga(CH_3)_3$ (TMG), which is the preferred metalorganic source of gallium for MOCVD. During the $Al_xGa_{1-x}As$ growth, $Al_2(CH_3)_6$ (TMA) was added to provide the Al source.

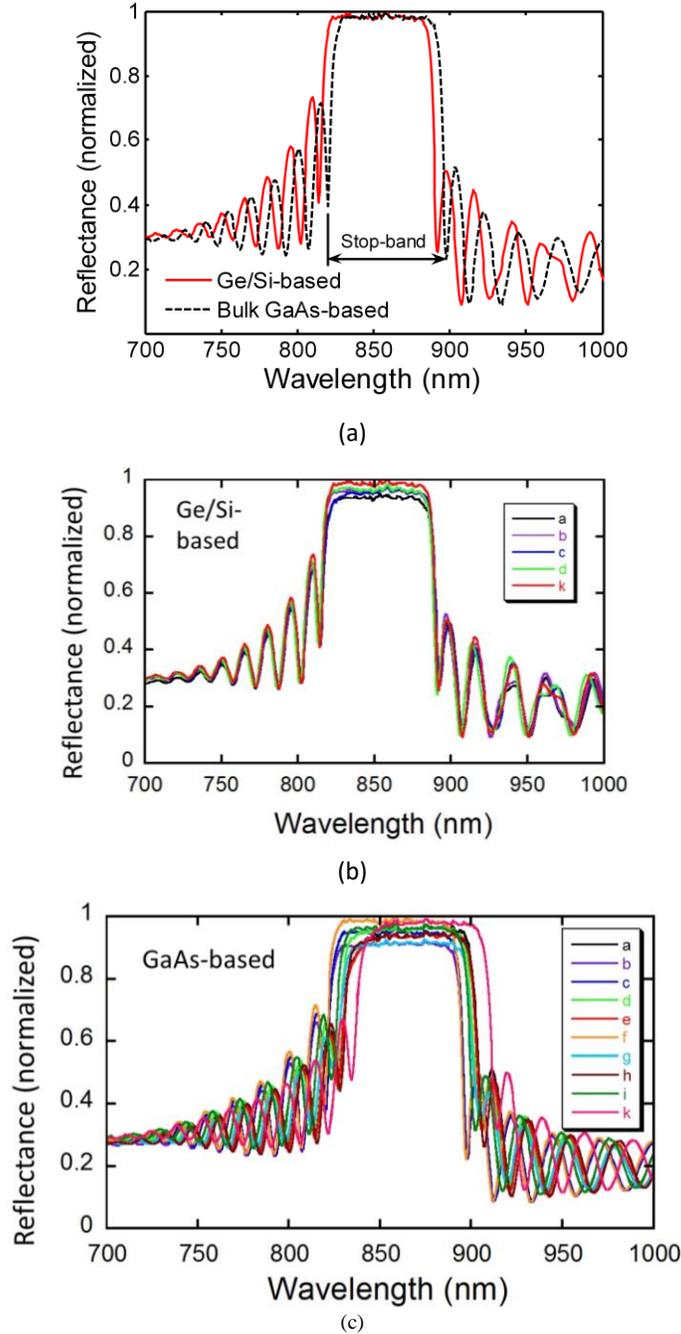

Fig. 2. (a) Normal-incidence reflectance spectra of bulk GaAs-based DBRs and ART-Ge/Si-based DBRs taken at the center of each sample normalized by the maximum intensity. The ART-Ge/Si-based DBRs were grown on parallel $SiO_2$ trench pattern with 1 μm trench spacing. (b) Normal-incidence reflectance spectra of the ART-Ge/Si-based DBRs measured at 5 spots (a to k) across a 12 mm by 12 mm region with the same trench pattern showing the cross-sample non-uniformity, which are normalized by the maximum intensity of the spectrum from the center spot "k" (the red line). (c) Normal-incidence reflectance spectra of bulk GaAs-based DBRs measured at 10 spots (a to k) across a 3" bulk GaAs wafer showing the cross-sample reflectance non-uniformity, which are normalized by the maximum intensity of the spectrum from the center spot "f" (the orange line).

In this work, because the Ge film was grown on on-axis (100) Si substrates, APDs become a primary concern due to the well-recognized polarity mismatch polar/nonpolar at the GaAs/Ge interface. Fortunately, the GaAs layer grown on the virtual Ge ART substrate was nearly free of APDs, measured previously in [15] by a scanning electron microscope. Growth calibration revealed that the tendency of APD formation on a virtual Ge ART substrate surface is different from that observed on a Ge (001) substrate. Firstly, it has been demonstrated that (113) facets are primarily formed during initial Ge growth inside the trenched area [18]. As the Ge film grows above the $SiO_2$ trenches, a coalesced wavy Ge growth front is formed, and faceting orientation gradually changes from (113) to (001). Therefore, in the vicinity of the coalesced Ge region, the as-polished Ge surface may have miss-oriented facets in a periodic style, which would provide an equivalent off-cut surface feature and lead to APD reduction in the overgrown GaAs layers. Secondly, we found that APDs can be further avoided using an optimized GaAs/AlGaAs superlattice structure before growing GaAs base layer on virtual Ge ART substrates.

## 3. Characterization results and discussions

Good DBR optical and material properties, i.e. reflectance spectra, smooth morphology and low threading dislocation density, are essential for the operation of VCSELs. To investigate the impact of the Ge/Si ART substrates on the DBR quality, optical imaging, high-resolution X-ray diffraction (HRXRD), cross-section transmission electron microscopy (XTEM) and etch pit density (EPD) analysis were used together with optical reflectance spectra measurements.

Fig. 2 (a) shows the normal-incidence reflectance spectra of the DBRs on the ART substrate and on the bulk GaAs substrate taken at the center of each sample, which were obtained at room temperature with a Filmmetrix F20 thin-film analyzer. The spectra of the GaAs/ART-Ge/Si-based DBRs have a very similar shape, stop-band width and peak height as those of the standard GaAs-based DBRs (control sample). The stop-band width of the GaAs/ART-Ge/Si-based DBRs and the bulk GaAs-based DBRs are 78.1 nm and 77.4 nm respectively. This difference between the Ge/Si-based DBRs and the bulk GaAs-based DBRs is within the cross-sample deviations of the bulk GaAs-based DBRs shown in Fig. 2 (c). The Ge/Si-based DBRs has a better cross-sample uniformity (Fig. 2 (b)) than that of the bulk GaAs-based DBRs, which may be a result of the smaller sample size of the former one. There is a noticeable spectral shift between the two samples. The stop-band centers of GaAs based-DBRs vary from 858 nm to 868 nm across the sample, while those of the GaAs/ART-Ge/Si-based DBRs range from 848 nm to 860 nm, closer to the 850 nm design target. Though the reflectance spectrum shows a pattern dependence of the ART oxide trench patterns, the stop-band centers of most patterns are shorter and closer to 850 nm (by 10 ~15 nm) than the DBR control sample. This implies the different properties between the Ge/Si ART substrates and bulk GaAs (e.g. thermal conductivity, offcut, air flow dynamics and size) have a small impact, if any, on the growth rate of the DBR superlattices.

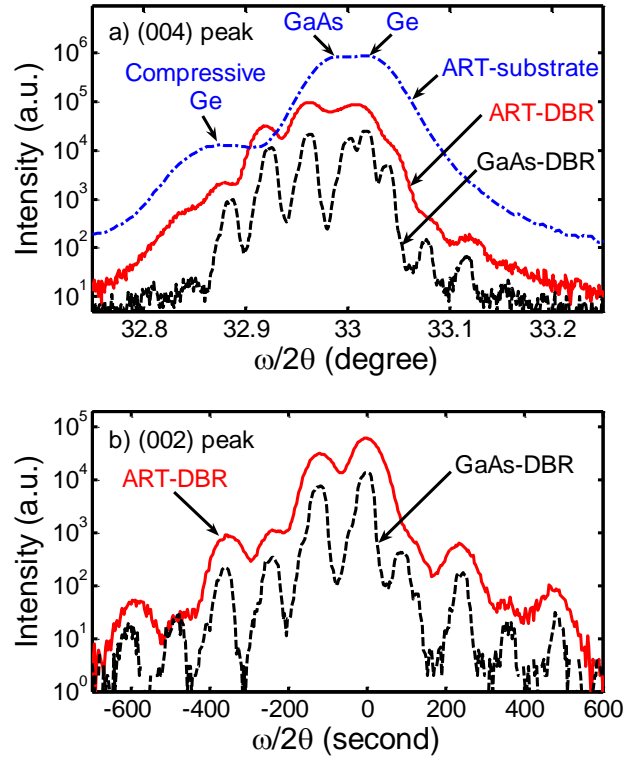

Fig. 3. (a) 004 X-ray scans of ART substrate, ART-based DBR (ART-DBR) and bulk GaAs-based DBR (GaAs-DBR); (b) 002 scans of ART-DBR and bulk GaAs-based DBR. The ART-substrate and ART-based DBRs were grown on parallel SiO2 trench pattern with 1 μm trench spacing.

In order to compare the crystal quality between the bulk GaAs-based and GaAs/ART-Ge/Si-based DBRs, HRXRD ω/2θ rocking curves were measured. We used Ge (220) 4-bounce monochromator for incident beams (accuracy = 0.004°) and a 1/4° slit for the detector. Fig. 3 (a) shows the (004) ω/2θ diffraction curves before and after DBR growth on ART substrates.

From the bottom to the top, the strain of Ge changes from in-plane compression (Ge in the trenches) to in-plane tension (the top portion of the coalesced Ge layer). In the ART-substrate XRD curve, the peak at 33.03º corresponds to the top of the coalesced Ge layer under +0.133% in-plane biaxial tensile strain and the peak at 32.99º is from the 2.5 μm GaAs buffer layers under -0.103% in-plane biaxial compressive strain, which shows that both GaAs and the coalesced Ge layer have been almost fully relaxed. The XRD curves of ART-Ge/Si were calibrated by the Si peaks from substrate. The lower intensity peak at 32.88º corresponds to the Ge in oxide trenches with a -0.425% in-plane biaxial compressive strain. These strains are superpositions of a very small residual epitaxy strain and a thermal mismatch strain, which was studied in a previous report [19], where Ge strain in oxide trenches and in the coalesced Ge film were measured separately and simulated. It is also observed that waffle patterns reduce the strain level for regions with the same oxide trench spacing.

The comparisons of crystal plane (004) and (002) ω/2θ diffraction between GaAs/ART-Ge/Si-based DBRs and GaAs-based DBRs are shown in Fig. 3 (a) and Fig. 3 (b) respectively. Though bulk GaAs-based DBRs have sharper satellite peaks, GaAs/ART-Ge/Si-based DBRs also show excellent structural quality, considering the effect of more complex substrate structure (e.g. $SiO_2$, Ge and strained-Ge). As X-rays penetrate GaAs, Ge, $SiO_2$ layers and the top of the Si substrates, a more complex substrate gives broader peaks. The superlattice period of bulk GaAs-based DBRs calculated from the DBR superlattice XRD fringes was 134.9 nm, slightly thicker than the designed value 129 nm.

XTEM samples were cut from ART-DBR samples by a focused ion beam (FIB) using an

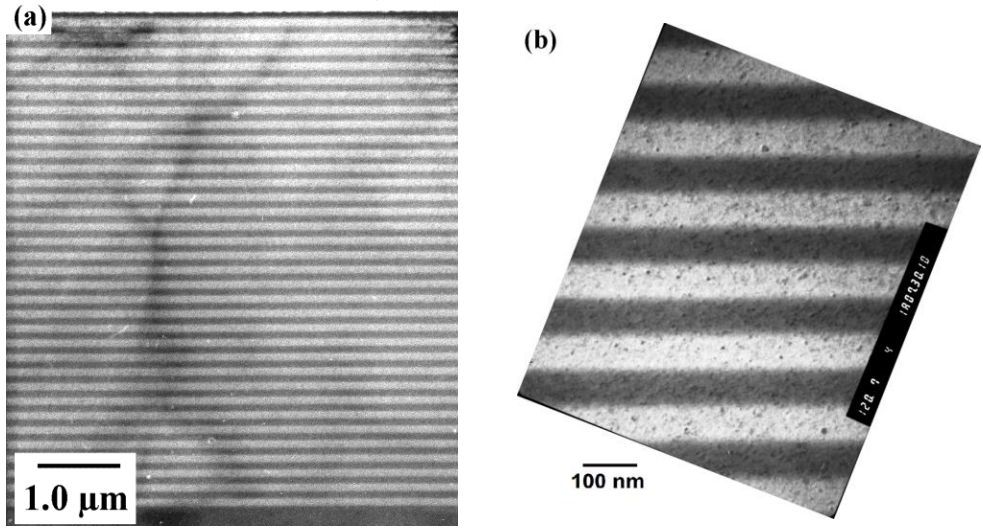

Fig. 4. Cross-section TEM images of Ge/Si ART-based DBRs at two different magnifications.

SMI 3050 two-beam system. The FIB resolution is 4 nm and the XTEM sample was less than 0.1 *μm* thick. Fig. 4 sample imaging was conducted using a Hitachi H-800 TEM. The operation voltage was 200 kV. XTEM images of GaAs/ART-Ge/Si-based DBRs in Fig. 4 (a) reveal reasonably good periodicity and uniformity. The shadowy lines are thickness fringes, which are TEM artifacts. A corresponding higher resolution image of the DBR in Fig. 4 (b) shows reasonably abrupt and smooth interfaces between the alternating layers. No threading dislocations are found within the TEM imaging range. A full structure image from the Si substrate to AlGaAs DBRs is shown in Fig. 5, which combines three XTEM images. Again, the wavy lines that run mainly from bottom to top are due to TEM sample thickness variations, which are not real features.

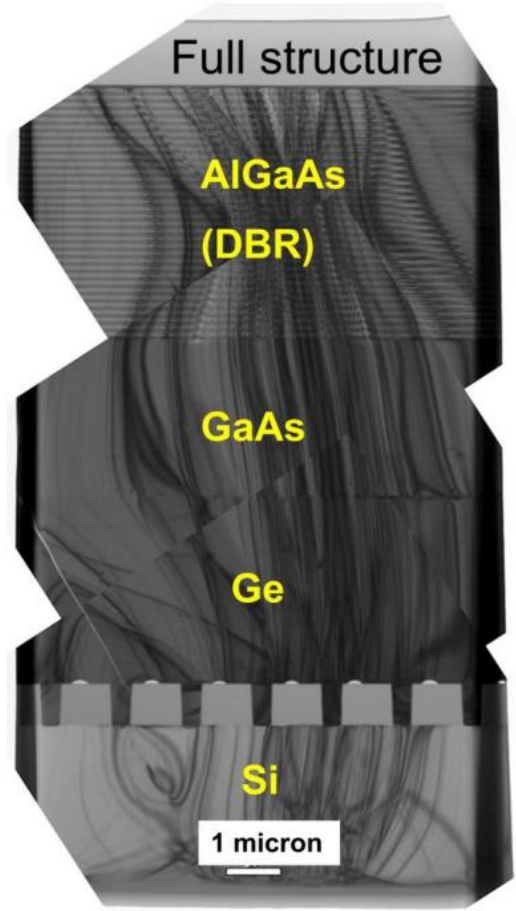

Fig.5. XTEM image of a full structure of GaAs/ART-Ge/Si-based DBRs.

EPD measurements were conducted on GaAs/Ge/Si ART substrates before the DBR growth to measure the threading dislocation density. An etch solution consisting of $CrO_3$, HF and $AgNO_3$ was used to etch the 2 μm GaAs base layers in the ART substrates [20]. EPD counted under a Nomarski microscope ranges from $10^5$ to $6 \times 10^6$ cm$^{-2}$ for different patterns, which agrees with a previous report [15]. Since those dislocations are populated in specific directions [13], it's possible to select relatively dislocation free regions for VCSEL active regions. Full VCSEL growth on GaAs/ART-Ge/Si-based substrates and the silicon oxide trench pattern impact on crystal quality and device performance are to be investigated next.

## 4. Conclusion

High quality AlGaAs DBRs were successfully monolithically grown on on-axis Si (100) substrates via a Ge layer formed by ART technique. The GaAs/ART-Ge/Si-based DBRs have reflectivity spectra comparable to those grown on conventional bulk GaAs substrates and have smooth morphology, and quite reasonable periodicity and uniformity. APD formation is significantly reduced in GaAs on ART-Ge/Si substrates, and etch pit density of the GaAs base layer on the ART-Ge substrates ranges from $10^5$ to $6 \times 10^6$ cm$^{-2}$. These results paved the way for future successful full VCSEL growth and fabrication on these ART-Ge substrates and also confirm that virtual Ge substrates via ART technique are effective Si platforms for optoelectronic integrated circuits.

## 5. Acknowledgment

This work was funded by Natural Science and Engineering Research Council of Canada (NSERC). The authors thank Dr. Huijun Huang at the Department of Materials and Optoelectronic Science, National Sun Yat-Sen University for her assistance in TEM imaging, and Guangnan Zhou at the Department of Materials Engineering, the University of British Columbia for his assistance in XRD data analysis.